\documentclass[preprint]{vgtc}               

\graphicspath{{figures/}{pictures/}{images/}{./}} 

\usepackage{times}                     

\usepackage{tabu}                      
\usepackage{booktabs}                  
\usepackage{lipsum}                    
\usepackage{mwe}                       
\usepackage{amsmath}

\usepackage{balance}

\usepackage{mathptmx}                  

\onlineid{6}

\vgtccategory{Research}

\vgtcinsertpkg

\preprinttext{Author's preprint version. To appear at IEEE ISMAR Workshops, 2026.}

\title{An AI-Driven Virtual Patient for Breaking Bad News: An Expert Formative Study on Facial Expression Intensity}

\author{Steffen Hauck\thanks{e-mail: steffen.hauck@stud.hs-coburg.de}\\ %
        \scriptsize Coburg University of Applied Sciences %
\and Theresa Schell\thanks{e-mail: theresa.schell@hs-coburg.de}\\ %
     \scriptsize Coburg University of Applied Sciences %
\and Sophie Jörg\thanks{e-mail: sophie.joerg@uni-bamberg.de}\\ %
     \scriptsize University of Bamberg %
\and Jens Grubert\thanks{e-mail: jens.grubert@hs-coburg.de}\\ %
     \scriptsize Coburg University of Applied Sciences}

\teaser{
  \centering
  \includegraphics[width=\linewidth]{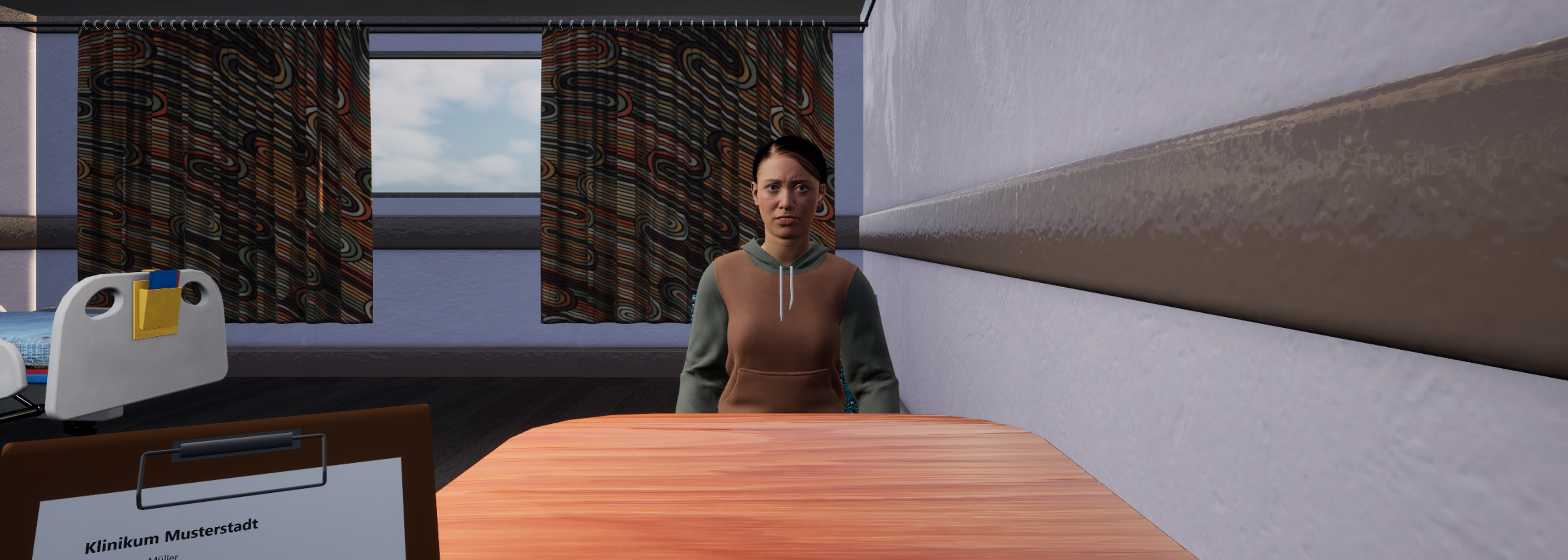}
  \caption{The VR training simulation setup: A consultation corner within a simulated hospital room, showcasing the face-to-face interaction with the AI-driven virtual patient during a cancer diagnosis communication.}
  \label{fig:AgentScreenshot}
}

\abstract{
    Interactive virtual patients driven by large language models (LLMs) offer scalable solutions for medical communication training, such as breaking bad news. However, designing their emotional expressiveness remains a challenge. This paper presents an AI-driven virtual patient framework combining LLM dialogue with real-time facial animation in virtual reality (VR). We conducted an exploratory, formative evaluation with seven medical experts to gather early feedback and elicit design requirements. The evaluation focused on how variations in facial expression intensity affect perceived realism and the virtual patient's emotion intelligibility. While descriptive quantitative ratings remained baseline across conditions, qualitative interviews provided deep insights into how experts perceive virtual emotional cues. The findings suggest that experts evaluate emotional realism holistically through multiple verbal and non-verbal channels; isolated facial adjustments are easily overshadowed by dialogue nuances and vocal prosody. Based on these insights, we present a roadmap for future medical training simulators, highlighting the need for synchronized, multi-modal pipelines incorporating body gestures, conversational pauses, and vocal dynamics.
} 


\begin{document}


\firstsection{Introduction}

\maketitle

Knowing how to deliver bad news is an integral part of being a doctor. Inadequate delivery of bad news can negatively affect patients' psychological well-being, trust in healthcare professionals, and their understanding of medical information \cite{Atasoy2012-dl} While historically overlooked, breaking bad news (BBN) has now become a cornerstone of modern medical education \cite{Fallowfield2004-wo}.
The gold standard for simulating these scenarios is working with standardized patients (actors). However, this approach is highly time- and resource-intensive, severely limiting doctors' access to frequent, flexible, and repeatable communication training \cite{bosse2015}. Virtual Reality (VR) combined with Embodied Conversational Agents (ECAs) offers a scalable and accessible alternative. While early systems relied on rigid, script-based virtual patients that only covered predefined paths \cite{dafli2015}, recent advances in generative AI allow for the creation of dynamic, unpredictable, and highly realistic dialogues from scratch. This enables a new paradigm of adaptive communication training through AI-driven virtual patients.
However, delivering bad news is inherently emotional, and words alone are insufficient to portray the gravity of the situation. Human-computer interaction research has demonstrated that the facial expressiveness of virtual agents in healthcare simulations is a critical factor in inducing empathy and emotional resonance in users \cite{milcent2022using}. Integrating such affective agents has been shown to encourage trainees to reflect on their own emotional responses and perceive the interaction as significantly more realistic \cite{loizou2024designing}. Crucially, recent evidence suggests that fostering such emotional resonance and empathy in medical students within VR environments can improve learning outcomes and enhanced knowledge retention \cite{xu2025enhancing}. \\
Building upon these findings, our approach is centered on the idea that manipulating the intensity of a virtual patient's emotional expressiveness can be used to trigger a stronger emotional reaction in physicians, thereby serving as a catalyst for training effectiveness. As a crucial first step toward validating this link, the non-verbal communication profiles must be carefully calibrated. In this paper, we present a modular VR prototype that integrates a generative AI-driven virtual patient capable of displaying different levels of emotional facial expression intensity. 
To evaluate this foundational component, we conducted a formative study with medical experts in a simulated breaking-bad-news scenario, comparing a low-intensity with a high-intensity expression condition regarding perceived realism and emotional intelligibility of the agent during this highly distressing conversation. Rather than presenting a finalized training system, the primary contribution of this work lies in deriving expert-informed design requirements for the next generation of emotionally expressive virtual patients in clinical training.

\section{Related Work}
Research on AI-driven virtual patients in medical education has shifted from rigid scripts toward dynamic, emotionally responsive simulations. This section reviews virtual patient systems tailored for BBN, the role of multi-modal expressiveness in virtual environments, and their impact on emotional resonance and training outcomes.
\subsection{Virtual Patients in Medical Breaking Bad News Training}
Communication training is a cornerstone of medical education, particularly for high-stakes clinical encounters such as delivering bad news. While traditional training relies on human actors as standardized patients \cite{bosse2015}, virtual patients have emerged as a highly scalable addition capable of helping develop communication skills \cite{carrard2020virtual}. Early systems were heavily restricted by predefined dialogue trees, failing to capture the unpredictable nature of real clinical communication. The integration of Large Language Models (LLMs) addresses this limitation by enabling open-ended, autonomous conversations that adapt dynamically to the user's input \cite{Steenstra2024alcoholCounseling}. Building upon these architectures, recent systems leverage LLMs to streamline dialogue authoring and support complex clinical communication techniques—such as motivational interviewing and health counseling role-play \cite{Nouraei2025HealthDial, Nouraei2025VirtualAgentSkills, Steenstra2025ScaffoldingEmpathy}. Within the specific context of BBN training, Ochs and colleagues developed the ACORFORMED platform, an interactive VR simulation system that allows doctors to practice delivering bad news to a virtual patient. This platform was utilized to collect multimodal interaction data \cite{ochs2018acorformed} and evaluate how different virtual displays affect users' sense of presence and believability of the virtual patient during the simulation \cite{ochs2019training}. While these studies establish the importance of environmental factors, integrating fully autonomous, generative AI dialogues with dynamic facial expressions remains an open area of exploration.
\subsection{Multi-Modal Expressiveness and Emotion Intelligibility}
In distressing medical encounters, human communication relies fundamentally on non-verbal signals. Consequently, incorporating expressive facial animation is critical to enhancing the social presence, perceived realism, and believability of virtual agents \cite{milcent2022using, rad2026designing, ashrafi2025emotionally}.
Empirical evaluations consistently highlight that specific non-verbal facial configurations—such as targeted sad or warm emotional expressions—are crucial for establishing user trust, perceived empathy, and believability across different demographics \cite{hosseinpanah2018empathy, parmar2022designing}.
Furthermore, immersive environments can further amplify this impact, as prior work in BBN training demonstrates that more immersive virtual displays positively influence the user's overall sense of presence and emotional engagement during the simulation \cite{ochs2019training}.
However, increasing expressiveness requires a careful balance: systematic investigations within health counseling domains reveal that excessive or poorly timed visual feedback can sometimes distract from persuasive health messages or interfere with the therapeutic bond \cite{parmar2022designing}. Moreover, studies investigating visual and emotional fidelity indicate that variations in an agent's emotional expression directly modulate user-perceived realism, though designers must carefully navigate emotional intensity to avoid adverse perceptual reactions \cite{higgins2023investigating}.
Furthermore, corpus analyses of clinical interactions highlight that emotional communication is inherently multi-modal, requiring a tight coordination of verbal and non-verbal channels \cite{ochs2018acorformed}. Despite these insights, relatively little research has systematically isolated how the visual intensity of facial expressions alone contributes to the perceived realism and emotion intelligibility of an agent when embedded in a fully autonomous dialogue system. Prior work in social VR suggests that subtle algorithmic modifications of nonverbal avatar behavior are not always consciously perceived by users. Roth et al. \cite{roth2018effects} manipulated nonverbal mimicry in real-time VR interactions and found that most participants did not detect the modification and that it did not significantly affect perceived communication quality. This supports the need to consider facial expression intensity as part of a broader multimodal behavior system rather than as an isolated visual parameter. However, investigating this isolated impact represents a critical baseline step, as it remains unclear whether systematic variations in facial expressiveness are sufficient to alter user perception in high-stakes conversations, or how they interact with the dominant verbal channels.
\subsection{Emotional Resonance and Training Effectiveness}
Beyond establishing immediate behavioral realism, the emotional expressiveness of a virtual patient directly drives the pedagogical value of the simulation. High levels of spatial and social presence in virtual environments have been shown to correlate positively with emotional engagement and overall learning retention \cite{lee2020effective}. Within the context of medical education, recent evidence demonstrates that fostering empathy and emotional resonance within VR simulations significantly improves learning outcomes for students \cite{xu2025enhancing}. Furthermore, experiencing authentic emotional distress from a virtual patient encourages learners to reflect on their own affective states and adopt more empathetic communication strategies \cite{carrard2020virtual}. However, while the pedagogical benefits of emotional engagement are well-established, it remains unclear which specific agent behaviors are required to trigger and sustain this resonance during fully autonomous dialogues. In particular, it remains under-explored whether adjusting the visual intensity of facial expressions alone is sufficient to influence training perceptions, or if a broader multi-modal pipeline is necessary. Our work addresses this gap through an exploratory evaluation of an AI-driven virtual patient with adjustable facial intensity, gathering expert requirements to guide the development of future medical training simulators.

\section{Prototype}
\begin{figure}[t]
    \centering
    \includegraphics[width=\columnwidth]{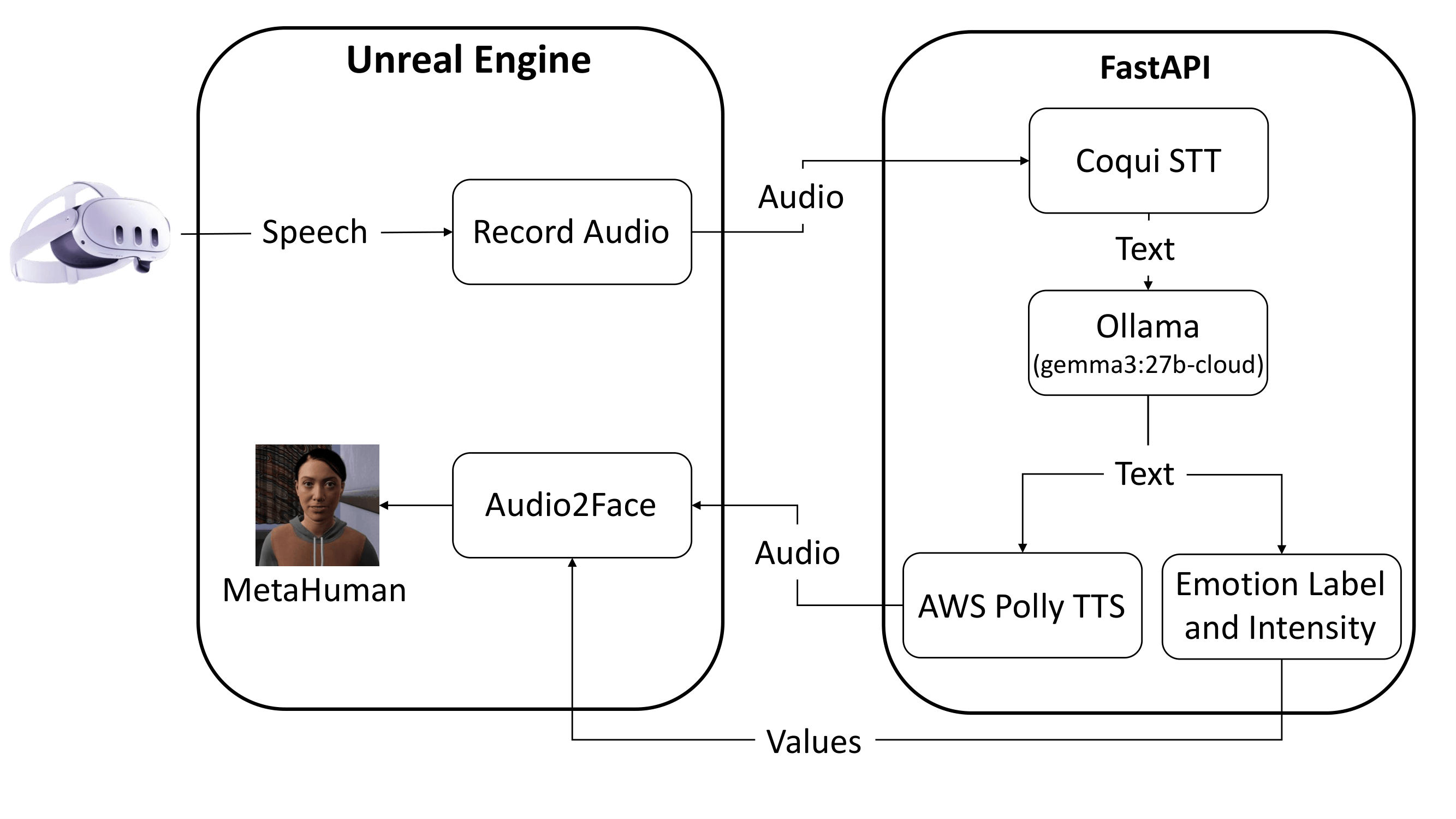}
    \caption{Overview of the proposed system architecture.}
    \label{fig:architecture}
\end{figure}

The primary objective of the prototype was to create an immersive VR experience that supports dynamic, AI-driven dialogue paired with context-aware emotional facial expressions. To ensure future-proofing in a fast-paced development landscape, the system builds upon a highly modular architecture (see Fig.~\ref{fig:architecture}), allowing individual AI services to be swapped as newer models emerge.
\subsection{System and XR Setup}
The front end was implemented in Unreal Engine 5.5.4. User speech is sent to a FastAPI backend for speech recognition, language generation, and speech synthesis. The generated audio and emotion label are then returned to Unreal Engine, where NVIDIA Audio2Face produces synchronized facial animations. The VR hardware setup consisted of a Meta Quest 3 head-mounted display (HMD) connected to a host PC via Meta Horizon Link. The PC hosting the simulation was equipped with an NVIDIA RTX A5000 GPU, 128 GB of RAM, and an Intel Xeon 1390P (3.5 GHz) CPU. Users interacted with the environment using the Quest 3 controllers, utilizing the grip and trigger buttons as the primary input mechanism to progress through the simulation. To minimize latency and optimize runtime performance on the local hardware, several core components of the pipeline were offloaded from local instances to high-performance cloud services.
\subsection{AI Integration and Dialogue Pipeline}
The conversational backbone of the prototype relies on an LLM-driven orchestration framework managed via a Python (3.10) backend server. While the system architecture natively supports Speech-to-Text (STT) input via a local Docker-hosted STT service, prompts were fed directly into the backend for the controlled conditions of this study to mitigate speech recognition variances.\\
Dialogue generation was handled via the Ollama cloud service running the gemma3:27b-cloud model. To ensure the virtual patient adhered to the medical context, the conversation was embedded into a predefined system prompt that defined the patient's persona, conversational boundaries, and required emotional output format. The LLM generates a text response accompanied by a specific emotion tag. For the final evaluation, Text-to-Speech (TTS) synthesis was handled via pre-baked high-fidelity voicelines generated with ElevenLabs to ensure natural prosody, replacing an initial AWS Amazon Polly cloud implementation.

\subsection{Facial Animation and Emotion Rendering}
To achieve realistic, runtime-generated facial expressions corresponding to unpredictable LLM outputs, pre-rendered animations were not viable. Instead, the system utilizes NVIDIA Audio2Face (A2F) via the NVIDIA ACE cloud service. Unreal Engine transmits the synthesized audio package alongside the emotion tag generated by the LLM (e.g., sadness, fear) to the A2F cloud interface. These emotion tags are parsed and dynamically translated into target blendshape configurations, which A2F blends in real time with the audio-driven lip-sync motion.
To operationalize different levels of emotional expressiveness, we expanded the data package sent to A2F by an intensity modifier manipulating the framework's internal "Overall Emotion Strength" parameter (range: 0.0--1.0). The specific intensity levels were established through iterative pilot testing during system calibration: values below 0.6 resulted in near-static facial geometry barely distinguishable from a neutral baseline, whereas 1.0 represented the maximum expressive boundary supported by A2F without introducing visual mesh distortions. Thus, 0.6 was chosen for the low-intensity condition to ensure subtle yet perceptible facial motion while maintaining a noticeable contrast to the 1.0 high-intensity setting. While A2F natively supports complex mixed-emotion rendering, this feature was bypassed to maintain distinct, controllable experimental conditions.
The complete data flow connecting these front-end rendering steps, backend orchestration, and cloud-based AI services is illustrated in Fig.~\ref{fig:architecture}.

\section{Evaluation}
To assess the viability of the proposed system and gather professional feedback on the simulation, we conducted an expert formative evaluation. This evaluation combined a controlled laboratory study with subsequent qualitative interviews.
\subsection{Study Design}
\begin{figure}[t]
    \centering
    \includegraphics[width=0.48\columnwidth]{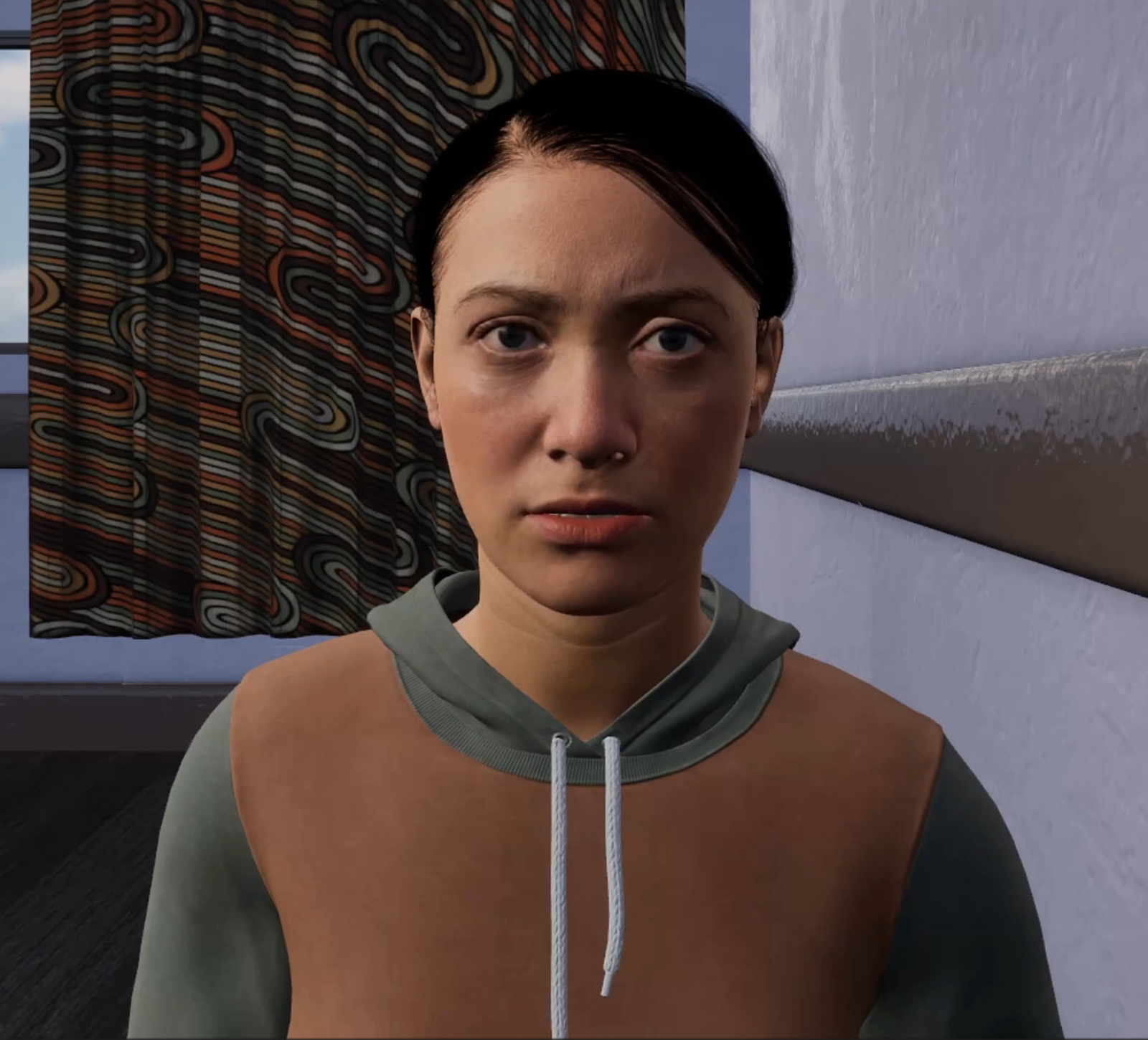}
    \hfill
    \includegraphics[width=0.48\columnwidth]{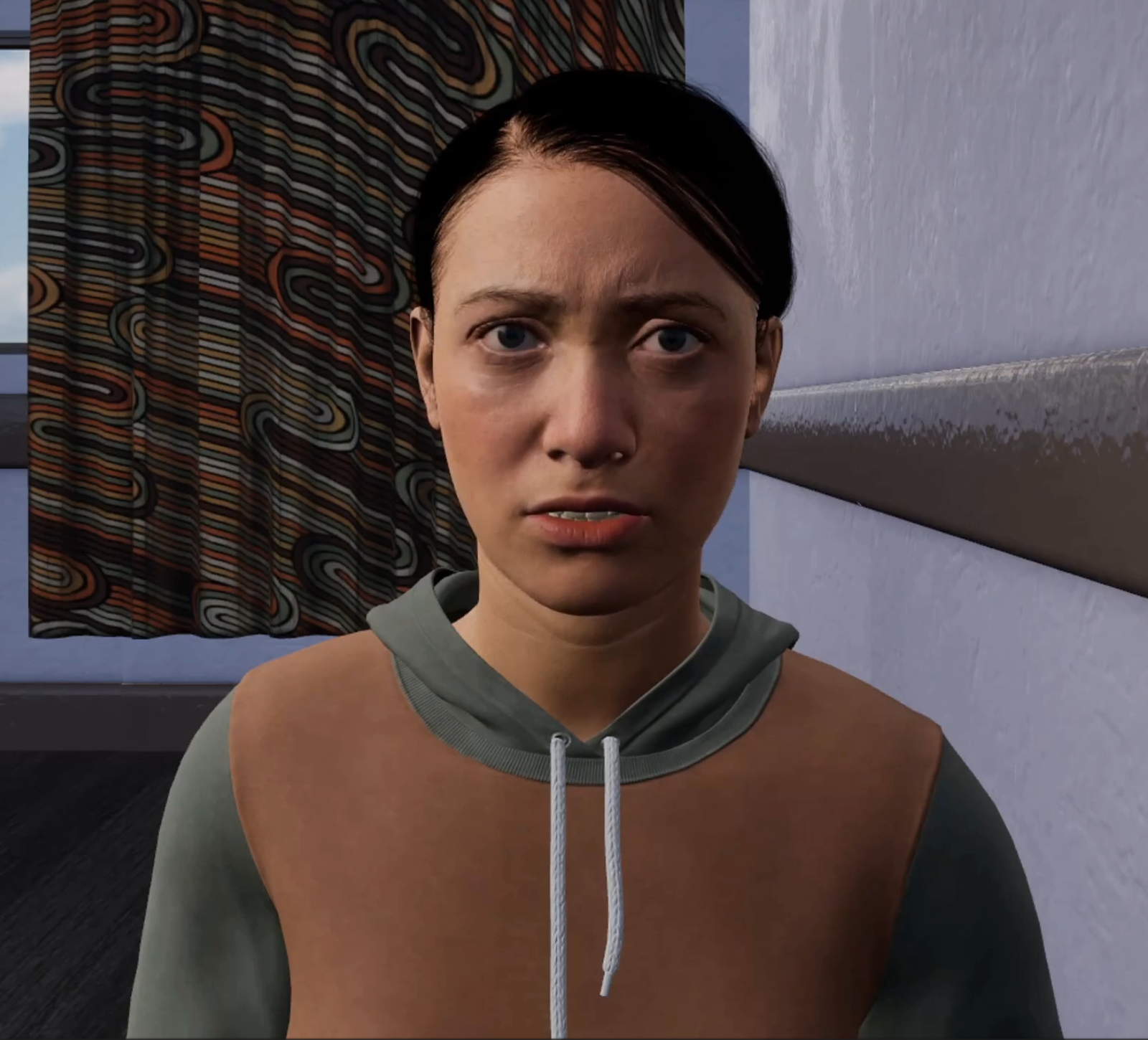}
    \caption{Comparison of the two emotional intensity conditions. Left: low intensity 0.6. Right: high intensity 1.0.}
    \label{fig:emotioncomparison}
\end{figure}

The investigation followed a within-subjects design with a single factor manipulated across two levels: facial expression intensity (low-intensity vs. high-intensity; see Fig.~\ref{fig:emotioncomparison}). Although the implemented architecture inherently supports fully dynamic, real-time AI-driven conversations, the evaluation employed two distinct, pre-generated dialogue scripts. Both scripts shared the identical medical context and diagnosis (lung cancer) to ensure a highly standardized emotional gravity, but utilized slightly varied phrasing and conversational flows to prevent learning effects or fatigue from repeating the exact same interaction. 
To ensure high ecological validity and clinical relevance, the dialogue structure in both variations was rigorously designed to follow the SPIKES protocol for BBN \cite{baile2000spikes}.

\subsection{Measurements}
We employed a mixed-methods approach combining quantitative psychometric questionnaires and qualitative interviews. The quantitative metrics comprised the 'realism' subscale of the Virtual Human Plausibility Questionnaire (VHPQ) \cite{mal2022virtual} to evaluate visual and behavioral fidelity, the 'emotion' subscale of the Virtual Agent Believability Scale (VABS) \cite{guo2023developing} to measure the intelligibility of the agent's affect, and the standard System Usability Scale (SUS) \cite{perrig2025development} to evaluate the technical usability of the VR setup.\\
The subsequent semi-structured closing interview gathered qualitative formative feedback, focusing on five core areas: 
\begin{enumerate}
    \item Consciously noticed differences between the two interaction conditions,
    \item Personal expectations and requirements for communication training simulations,
    \item The appropriateness and sensitivity of using a cancer diagnosis as the underlying medical scenario,
    \item The perceived utility of the system for medical students from an expert's perspective, and
    \item General feedback, criticisms, or suggestions for system improvement.
\end{enumerate}

\subsection{Task and Procedure}

Participants were tasked with experiencing a simulated clinical encounter where a virtual patient receives a severe diagnosis, see Fig.~\ref{fig:AgentScreenshot}. The simulation served as a demonstrator of how an interaction within the proposed VR training environment unfolds. Participants could progress through the structured dialogue at their own pace using the trigger or grip buttons on the VR controllers. Pressing the button initiated a pre-recorded audio playback of the clinician's dialogue line, which was then followed by the corresponding response from the virtual patient. \\
The standardized experimental procedure lasted approximately 30 minutes in total and was structured as follows: First, participants received a brief introduction to the system and the scope of the study, followed by the signing of an informed consent form and the completion of a demographic questionnaire. Participants then entered the first VR scenario (either low or high intensity, counterbalanced to avoid order effects). Immediately after exiting the first scenario, they completed the VHPQ and VABS questionnaires. Next, participants experienced the second VR scenario featuring the remaining intensity level, followed by the same set of post-scenario questionnaires. Finally, participants completed the SUS to assess the overall system, followed by a semi-structured closing interview.

\subsection{Participants}
A total of $N = 7$ medical experts (4 female, 3 male) participated in the study. The sample consisted of experienced clinicians recruited from specialized medical departments, including Oncology ($n = 2$), Gynecology ($n = 3$), Emergency care ($n=1$), and a specialized Pulmonology Center ($n = 1$). The participants' ages ranged from 27 to 66 years ($M = 47.14$, $SD = 12.55$). Reflecting their expert status, their professional clinical work experience ranged from 1 to 45 years ($M = 21.00$, $SD = 13.38$). Baseline experiences regarding communication and technology were assessed using 7-point Likert scales ($1 = \text{no experience}$, $7 = \text{expert}$). The clinical expertise is further underscored by a high baseline familiarity with breaking bad news ($M = 6.00$, $SD = 0.82$). Conversely, the participants reported lower prior exposure to virtual reality ($M = 2.00$, $SD = 1.29$) and moderate experience with simulation-based training modalities ($M = 2.86$, $SD = 1.21$) (see Table~\ref{tab:demographics}).

\begin{table}[tb]
  \caption{Demographic characteristics and baseline experience of the participants ($N = 7$), including the mean, standard deviation (in parentheses), and value range [in brackets].}
  \label{tab:demographics}
  \scriptsize%
  \centering%
  \renewcommand{\arraystretch}{1.3}
  \begin{tabular*}{\columnwidth}{@{\extracolsep{\fill}}llll@{}}
  \toprule
  \textbf{Characteristic / Variable} & & \textbf{Mean (SD)} & \textbf{[Min, Max]} \\
  \midrule
  \multicolumn{4}{l}{\textit{Demographics}} \\
  \hspace*{3mm}Age (years) & & 47.14 (12.55) & [27, 66] \\
  \hspace*{3mm}Professional Experience (years) & & 21.00 (13.38) & [1, 45] \\
  \midrule
  \multicolumn{4}{l}{\textit{Prior Experience (7-point Likert Scale)}} \\
  \hspace*{3mm}Breaking Bad News & & 6.00 (0.82) & [5, 7] \\
  \hspace*{3mm}Virtual Reality & & 2.00 (1.29) & [1, 4] \\
  \hspace*{3mm}Simulation-Based Training & & 2.86 (1.21) & [2, 5] \\
  \bottomrule
  \end{tabular*}
\end{table}

\section{Results}
\label{sec:results}

\begin{table}[tb]
  \caption{Descriptive statistics: Mean and standard deviation (in parentheses) for dependent variables. VHPQ and VABS subscale scores are shown per condition (low vs. high intensity), while SUS represents the overall global evaluation post-exposure.}
  \label{tab:descriptiveStatistics}
  \scriptsize%
  \centering%
  \renewcommand{\arraystretch}{1.3}
  \begin{tabular*}{\columnwidth}{@{\extracolsep{\fill}}lcc@{}}
  \toprule
  \textbf{Questionnaire / Measure} & \textbf{Low} & \textbf{High} \\
  \midrule
  \textit{VHPQ (Realism Subscale)} & 5.26 (0.66) & 5.26 (0.69) \\
  \midrule
  \textit{VABS (Emotion Subscale)} & 4.82 (1.54) & 5.11 (1.24) \\
  \midrule
  \midrule
  \textit{SUS (Usability)} & \multicolumn{2}{c}{76.07 (12.82)} \\
  \bottomrule
  \end{tabular*}
\end{table}

Given the exploratory nature of this formative study and the small sample size ($N=7$), quantitative results are reported descriptively and complemented by qualitative expert feedback.
\subsection{Quantitative Results}
The perceived realism subscale of the VHPQ remained descriptively identical across both experimental conditions (Low: $M=5.26$, $SD=0.66$; High: $M=5.26$, $SD=0.69$). Similarly, the emotional believability subscale of the VABS showed only a minimal descriptive increase for the high mimic intensity condition ($M=5.11$, $SD=1.24$) compared to the low intensity condition ($M=4.82$, $SD=1.54$). 
Overall technical usability of the prototype was rated positively, resulting in a mean SUS score of $76.07$ ($SD=12.82$), which corresponds to an "above-average" or "good" level of usability. The descriptive statistics for all the dependent variables are presented in Table ~\ref{tab:descriptiveStatistics}.

\subsection{Qualitative Feedback}
\label{sec:qualitative-feedback}
To gain deeper insights into these nuances, the qualitative interview data were analyzed using qualitative content analysis following Mayring \cite{mayring2021qualitative}. Four core themes emerged from the evaluation: (1) acceptance of the VR training concept, (2) perception of emotional communication, (3) clinical suitability of the selected scenario, and (4) concrete suggestions for future system improvements.

\subsubsection{Acceptance of the Training Concept}
All seven participants explicitly stated that they would have highly welcomed the integration of such a system during their own medical education. Several experts highlighted the immense flexibility of an AI-driven, VR-based solution compared to traditional standardized patients, emphasizing that it allows for highly realistic communication training independent of time, personnel, and organizational constraints. However, multiple participants emphasized that such an autonomous system should complement rather than replace expert human supervision. Combining self-directed VR training sessions with subsequent structured feedback from experienced physicians was considered the most promising pedagogical approach.

\subsubsection{Perception of Emotional Communication}
Regarding the emotional presentation of the virtual patient, the qualitative feedback provided a clear explanation for the overlapping quantitative scores: participants primarily perceived emotional shifts through verbal communication and vocal prosody rather than facial expressions. Several experts explained that subtle changes in wording, voice modulation, and the emotional gravity of individual statements made one of the scenarios appear more emotional, whereas the visual manipulation itself went largely unnoticed. In fact, only a single participant reported consciously noticing differences in the agent's facial expressions between both scenarios. Conversely, four participants independently noted that body gestures and posture would contribute substantially more to the perceived realism of emotionally challenging conversations than facial mimicry alone. One participant summarized this lack of non-verbal dynamics by stating that, apart from blinking, the virtual patient displayed "no facial expression and no body language."

\subsubsection{Suitability of the Selected Scenario}
The simulated lung cancer diagnosis was universally considered an appropriate and highly authentic scenario for communication training. Participants argued that a cancer diagnosis carries an immediate, profound emotional significance that is instantly understood by both medical professionals and patients. Furthermore, experts emphasized that such a diagnosis drastically affects multiple dimensions of a patient's life—including family dynamics, professional career, and long-term future planning. This multi-layered impact makes it an ideal framework for training complex protocols like SPIKES \cite{baile2000spikes}.

\subsubsection{Suggestions for Future Improvements}
Finally, the experts outlined several critical avenues for iterative advancements of the prototype. The most prominent recommendation was the urgent integration of dynamic body gestures to align with the vocal output. Furthermore, experts recommended expanding the emotional repertoire of the virtual patient to support severe, sudden stress responses such as crying, prolonged silences, physical agitation, or acute shock. Lastly, several participants remarked that the current strict question-answer cadence felt less natural than actual clinical encounters, advising that future versions should allow for conversational pauses, overlapping speech, interruptions, and spontaneous emotional outbursts.

\section{Discussion}
The primary objective of this formative evaluation was to exploratively assess how variations in the facial expression intensity of a virtual patient impact perceived realism and emotion intelligibility within a medical BBN scenario. While the descriptive quantitative results remained nearly baseline across both experimental conditions, the qualitative insights gathered from the medical experts provide profound guidance for the future development of emotionally expressive virtual patients.\\
A key revelation from the evaluation is that the systematically manipulated visual differences in facial expression intensity had a negligible impact on overall perception. The qualitative feedback provides two highly constructive explanations for this result, which offer a valuable foundation for subsequent research. \\
First, the experts evaluated the clinical encounter holistically as a multi-modal experience, predominantly focusing on dialogue nuances and vocal prosody rather than facial expressions in isolation. Subtle variations in phrasing and voice modulation between the two scenarios were perceived as emotional drivers, whereas the visual facial adjustments went largely unnoticed. This echoes findings by Roth et al. \cite{roth2018effects}, which similarly demonstrate that subtle, isolated modifications of nonverbal behavior often fail to elicit consciously perceivable differences when embedded in real-time interactions.\\
Second, the interviews suggest that the operationalized contrast between the low- and high-intensity settings within the facial animation pipeline may have been too subtle to register under the high cognitive and emotional load of a cancer diagnosis simulation. This critical finding indicates that future iterations must implement more pronounced visual thresholds and broader emotional dynamics to achieve a perceptually distinct manipulation.\\
Furthermore, the experts strongly emphasized that facial expression alone is insufficient for authentic patient simulation. In highly distressing scenarios, non-verbal cues such as dynamic body gestures, posture shifts, and systemic stress responses (e.g., weeping, freezing, or profound conversational pauses) are perceived as significantly more critical to realism than microscopic facial adjustments. Consequently, future communication systems must model emotional behavior holistically across multiple synchronous channels instead of focusing exclusively on facial pipelines. \\
Importantly, despite the visual subtleties, the evaluation demonstrated an exceptionally high acceptance of the overall generative AI training concept. The agreement among all expert participants that they would have highly valued access to such a system during their medical education underscores a strong, unfulfilled pedagogical demand. This validation confirms that combining LLMs for fluid dialogue with real-time animation addresses a core educational need, while our findings precisely map out the roadmap required to mature these systems.

\subsection{Limitations and Future Roadmap}
Several limitations of this formative study should be highlighted as guiding vectors for future research. Given the highly exploratory nature and the small sample size of medical experts ($N=7$), our quantitative data serve purely descriptive purposes and do not allow for broader statistical generalizations. \\ 
Furthermore, to maintain experimental control and prevent learning effects, the evaluation relied on two slightly varied, pre-generated dialogue scripts instead of a fully autonomous, real-time conversational loop. While necessary at this stage to isolate the specific variables, this constraint temporarily restricted the fluid interactivity that constitutes a primary advantage of LLM-driven virtual humans. Finally, the slight textual and prosodic variations between the two scripts introduced a confounding factor, meaning the visual facial manipulation was not fully isolated from verbal and vocal cues, as qualitative feedback indicated that text and prosody dominated visual perception.\\
Moving forward, these insights establish a roadmap for VR-based medical simulators. Our immediate next step will focus on expanding the virtual patient's capabilities towards an integrated, multi-modal animation pipeline that synthesizes automated body gestures and posture changes alongside facial expressions. Additionally, future iterations will implement higher visual intensity thresholds and transition to fully autonomous, real-time conversations. Ultimately, we aim to deploy the refined prototype in a broader evaluation with medical students to investigate how the synergetic interplay of multi-modal emotional expressiveness impacts overall training effectiveness in medical education.

\section{Conclusion}
This paper presented an AI-driven virtual patient designed for VR-based communication training in BBN scenarios. The system combines LLM-driven dialogue with real-time facial animation to enable natural interactions. Our formative evaluation with seven medical experts focused on exploratively assessing how variations in facial expression intensity influence the perceived realism and emotional intelligibility of the virtual patient, thereby establishing design requirements for future emotionally expressive agents. 

Although the quantitative differences in facial expression intensity did not lead to measurable changes in perception, the subsequent qualitative feedback provided valuable practical insights for future development. The experts' responses indicate that the realism of such interactions is evaluated based on a holistic combination of multiple verbal and non-verbal channels rather than facial animation alone. For future systems, this highlights the necessity of integrating broader communication modalities, such as body gestures and vocal nuances. Despite the subtle effects of the visual manipulation, the overall positive feedback from the participants confirms a strong interest in using AI-driven virtual patients as a valuable supplement to traditional medical education.

\section{Acknowledgements}
Generative AI (ChatGPT-5, OpenAI) was used to assist with language editing and to improve the flow and readability of the manuscript. The authors reviewed and assume full responsibility for the content of this article.

\balance

\bibliographystyle{abbrv-doi}

\bibliography{template}
\end{document}